\documentstyle[twoside,fleqn,espcrc2,epsf]{article}

\newcommand{\AmS}{{\protect\the\textfont2
  A\kern-.1667em\lower.5ex\hbox{M}\kern-.125emS}}

\title{Probing The CORE of the Haldane Conjecture}
\author{Marvin Weinstein \address{%
	Stanford Linear Accelerator  Center \\
	Stanford University, Stanford, California 94309 }%
        \thanks{This work was supported by the U.~S.~DOE, Contract
		No.~DE-AC03-76SF00515 .}}

\begin{document}
\begin{abstract}
The COntractor REnormalization group method (CORE), originally developed for application to lattice gauge theories, is very well adapted the study of spin systems and systems with fermions.  As an warmup exercise for studying Hubbard models this method is applied to spin-1/2 and spin-1 anti-ferromagnets in one space dimension in order to see if it is able to explain the physics of the Haldane conjecture.  The method not only provides support for Haldane's conjecture but provides insight into the physics of a more general class of spin-1 systems with Hamiltonians of the form $H= \sum_j \vec{s}(j)\cdot\vec{s}(j+1) - \beta\,( \vec{s}(j)\cdot\vec{s}(j+1) )^2$ about which, until now, little was known.
\end{abstract}

\maketitle

\section{INTRODUCTION}

Sometimes, when carrying out a preliminary computation in order to be sure that well understood llimits of a problem are handled correctly,  results which are interesting in their own right emerge. I will  now discuss one such result which I have obtained as part of such a preliminary study  of the Hubbard model.

The Hubbard model at half-filling is equivalent to a Heisenberg anti-ferromagnet (HAF) and it is important to show that CORE treats this limiit correctly.  This was shown to be the case for the spin-1/2 HAF in Ref.\protect\cite{COREpaper}.  What that paper did not address was the subtle question of how the physics of the one-dimensional anti-ferromagnet changes when the spin-1/2 on each lattice site is replaced by spin $S$.  In 1983 it was argued by F.D.M.~Haldane, Ref.\protect\cite{Haldane} that when $S$ is a half-integer, then the spectrum has no mass-gap, but when $S$ is an integer, a mass-gap develops.  In this talk I describe a CORE computation for the spin-1/2 and spin-1 anti-ferromagnet which not only supports Haldane's conjecture, but shows how the spin-1 case leads to an understanding of a more general class of theories defined by a Hamiltonian of the form
\begin{equation}
	H =  \sum_j   \vec{s}_j\cdot\vec{s}_j - \beta\,\sum_j (\vec{s}_j\cdot\vec{s}_j)^2 	 	   
\label{betaham}
\end{equation}   
about which, until now, very little was known.  . 

\subsection{What Is CORE?}

The CORE method consists of two parts.  First, a theorem which defines the Hamiltonian analogue of Wilson's exact renormalization group transformation;  second, a set of approximation procedures which render the calculation of the {\it renormalized Hamiltonian\/} doable.  CORE replaces the Lagrangian notion of integrating out degrees of freedom  by that of throwing away Hilbert space states;  i.e., defining a ojection operator, $P$, which acts on the original Hilbert space,   ${\cal H}$ to create the space of {\it retained states\/}  ${\cal H}_{\rm ret} = P {\cal H}$.   The central feature of this approach is a formula which relates the original Hamiltonian, $H$, to a {\it renormalized Hamiltonian\/}, $H_{\rm ren}$, which acts on ${\cal H}_{\rm ret}$ and which has, in a sense which was made precise in Ref.\protect\cite{COREpaper},  exactly the same low energy physics as $H$. This formula is 
\begin{equation}
        H^{\rm ren}= \lim_{t\rightarrow \infty} 
            \sqrt{[[T(t)^2]]} [[T(t) H T(t)]] \sqrt{[[T(t)^2]]} 
\label{basicform}
\end{equation}
where $T(t) = e^{-t H}$ and where we have defined $[[O]]= P O P$ for any operator $O$ which acts on ${\cal H}$.  

\subsection{Cluster Decomposition}

It was explained in Ref.\protect\cite{COREpaper} that $H^{\rm ren}$ is an extensive operator and can be written as a sum of finite-range connected operators; i.e.,
 \begin{equation}
        H^{\rm ren} = \sum_{j=-\infty}^{\infty} \sum_{r=1}^{\infty}
               h(j,r)^{\rm conn}
\label{cluster}
\end{equation}
where each $h(j ,r)^{\rm conn}$ can be computed on a finite sublattice of the original infinite lattice.   Although this formula involves an infinite number of terms examples have shown that a highly accurate approximation to $H^{\rm ren}$ only requires computing a few terms in the cluster expansion, typically range-2 and/or range-3 terms are adequate. 
            
\section{CORE: THE SPIN-1/2 HAF}

To obtain the Hamiltonian for a general Heisenberg anti-ferromagnet set $\beta=0$ in Eq.\ref{betaham}.  The complete discusion of a range-2 calculation for the spin-1/2 HAF appears in Ref.\protect\cite{COREpaper} and only a brief review of the results will be discussed here.  All CORE computations begin with a choice of a thinning procedure and the procedure used in Ref.\protect\cite{COREpaper} was based upon dividing the lattice into three site blocks $B_p$,  and solving the corresponding block-Hamiltonian (obtained by restricting the full Hamiltonian to those terms which are completely contained in block $B_p$), keeping a small set of low-lying eigenstates.  Since the HAF Hamiltonian has an $SU(2)$ symmetry its eigenstates fall into degenerate multiplets and to guarantee that the renormalized Hamiltonian has the same symmetry as the original  it is necessary to keep full multiplets.

Three-site blocks were chosen for the spin-1/2 HAF because that is the smallest size block for which a non-trivial CORE computation can be done.  This is because the eigenstates of the  two-site block-Hamiltonian fall into spin-0 and spin-1 multiplets, with the spin-0 state having the lowest energy and keeping only the lowest lying multiplet would amount to keeping a single state, a trivial renormalization group transformation.   Going to three sites avoids this problem since the eigenstates of the block Hamiltonian fall into two  spin-1/2 multiplets and one spin-3/2 multiplet.  In this case one of the spin-1/2 multiplets has the lowest energy and so truncating to the space generated by keeping only these two-states per block leads to a renormalized Hamiltonian for which the operator $h(p,1)^{\rm conn}$ is a multiple of the unit matrix and $h(p,2)^{\rm conn}$ is, up to a multiplicative constant which is less than unity, exactly the same starting Hamiltonian; i.e., the spin-1/2 HAF Hamiltonian is at a fixed point of the CORE transformation.  Iterating this calculation shows the spin-1/2 HAF is massless, in agreement with what is known about this system.  The same calculation produces a ground-state energy density which is within one-percent of the exact answer.

\subsection{CORE: The Spin-1 Case}

Applying the same three site blocking to the spin-1 HAF yields a different result.  In this case the eigenstates of the block Hamiltonian fall into a spin-0, spin-1 and spin-2 multiplet, with the spin-1 multiplet having the lowest energy.  Truncating to the space generated by keeping only the spin-1 multiplet per block gives, as in the spin-1/2 case, $h(p,1)^{\rm conn}$ proportional to the unit matrix, but now $h(p,2)^{\rm conn}$ takes the more general form
\begin{eqnarray}
h(p,2)^{\rm conn} &=& \vec{s}(p)\cdot \vec{s}([p+1) \nonumber\\
	&  & \qquad - \beta \left(\vec{s}(p)\cdot \vec{s}([p+1)\right)^2 \nonumber\\
\end{eqnarray}
Thus, after one three site blocking the CORE analysis of the spn-1 HAF turns into the study of renormalization group flows for the theory defined by 
\begin{equation} 
H = \sum_j	h(j,2)^{\rm conn}
\label{genham}
\end{equation}
where I have ignored the term proportional to the identiy operator.  Except for the special values $\beta=-1,-1/3,1$ very little is known about the Hamiltonian Eq.\ref{genham}.   The case $\beta=-1/3$ is special, in that it can be solved for the groundstate, which is a valence-bond solid with a non-vanishing mass gap.  Moreover, when $\beta=1$ the theory can be solved by means of the Bethe-ansatz and is massless.  Finally, when  $\beta = -1$ the theory becomes $SU(3)$ symmetric and it is conjectured that the mass gap vanishes at this point too, although no exact result is known for this case.      

Unlike the spin-1/2 case sensible CORE computations can be done for the spin-1 system defined by Eq.\ref{genham} using either two or three-site blocking.  If two-site blocks are used, then the states of the block Hamiltonian fall into a spin-0, spin-1 and spin-2 representation and keeping the spin-0 and spin-1 states defines a non-trivial renormalization group transformation.  The determination that the spin-0 and spin-1 multiplets are the ones to keep is made by studying the $\beta$ dependence of their energies.  This shows that one of these representations always lies lowest in energy and that they become degenerate and then cros at $\beta=-1/3$.  Similarly, the eigenstates of the three-site Hamiltonian divide into one spin-3, two spin-2, three spin-1 and one spin-0 representations of $SU(2)$ and a study of the $\beta$ dependence of the eigenenergies shows that the lowest lying spin-0 and spin-1 states are the ones to keep because one or the other of these multiplets always lies lowest in energy and they become degenerate and cross at $\beta=-1/3$.  Since either two or three site blocking leads to the same truncation algorithm I study the two-site blocking procedure because it is easier to carry out.

The approximate, range-2, two-site blocking CORE computation of the mass gap for  $-1  \le \beta  < 1.3$ is shown in the figure.  Several points should be made about this plot.  First, the CORE computation solves the $\beta=-1/3$ case exactly.   Analysis of the flow at this point shows that at each renormalized Hamiltonian is isomorphic to a dimerized spin-1/2 anti-ferromagnet and the fixed point is the place where one of the two nearest neighbor couplings vanish.  Furthermore, all of the Hamiltonians defined by Eq.\ref{genham} flow to the same massive fixed point for the range of $\beta$ shown in the graph.  The general picture which emerges from this calculation is that there are three fixed points to the flow in this region, an attractive, massive fixed point at $\beta=-1/3$ (the valence-bond solid) and two repulsive massles fixed points near $\beta=\pm 1$. The fact that the mass gap doesn't vanish for $\beta= \pm 1$ but for a value of $\beta$ which is 10\%  further from the valence bond solid point is typical of what occurs in range-2 approximations.  Going to range-3 or better should reduce this error by an order of magnitude.   The reason I cut off the plot at $\beta=-1$, before the massgap goes to zero, is that  at this point the theory becomes $SU(3)$ symmetric (a fact which is obvious from the CORE computation) and so the spin-0 and spin-2 representations become degenerate.  Because the program I wrote to do this calculation just grabbed the four lowest lying states the calculation destroys the manifest $SU(2)$ symmetry.a  Clearly a more careful programming job has to be done to go beyond this pont keeping the symmetry intact.  Note, the calculated mass gap is shown by the solid curve, the dotted curve is just included to heuristically remind the reader of what might be expected in an exact calculation.  Note, the spin-1 HAF at $\beta=0$ lies quite close to the $\beta=-1/3$ point where the CORE computation is exact and is well within in the basin of attraction of the massive theory, thus providing the advertised support for the Haldane conjecture.

\begin{figure}[htbp]
\epsfxsize 2.5in
\epsfbox{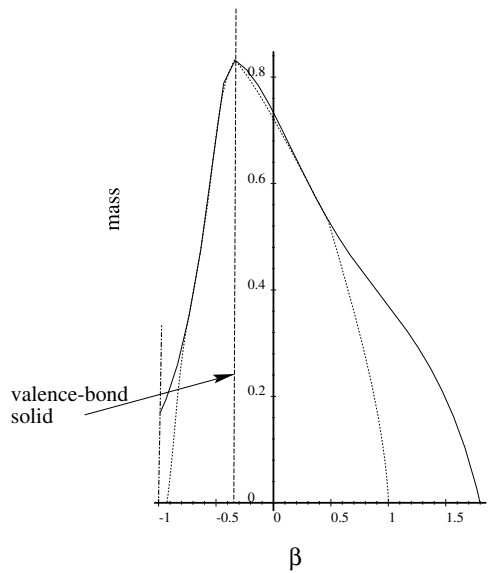}
\caption{Mass gap versus $\beta$ for range-2 computation}
\end{figure}

\end{document}